\documentclass[11pt,aps,pra,reprint]{revtex4-2}          % twocolumn

\RequirePackage[T1]{fontenc}

\RequirePackage{graphicx}
\RequirePackage{mathptmx}      % use Times fonts if available on your TeX system
\RequirePackage{flushend}
\RequirePackage{natbib}
\setcitestyle{numbers,sort&compress}
\RequirePackage[colorlinks,citecolor=blue,urlcolor=blue,linkcolor=blue]{hyperref}
\RequirePackage{amsmath}
\RequirePackage{tabularx}
\RequirePackage{dcolumn}
  \newcolumntype{d}[1]{D{.}{.}{#1}}
\RequirePackage{booktabs}
\RequirePackage[shortcuts]{extdash}
\begin{document}

\title{Competing Ionization and Dissociation in the H$_2$ Gerade System}

\author{D\'{a}vid Hvizdo\v{s}}
\email{david.hvizdos@jh-inst.cas.cz}
\affiliation{J. Heyrovsk\'{y} Institute of Physical Chemistry, ASCR, Dolej\v{s}kova 3, 
            18223 Prague, Czech Republic}
\author{Roman \v{C}ur\'{\i}k}
\email{roman.curik@jh-inst.cas.cz}
\affiliation{J. Heyrovsk\'{y} Institute of Physical Chemistry, ASCR, Dolej\v{s}kova 3, 
            18223 Prague, Czech Republic}
\author{Chris H.~Greene}
\email{chgreene@purdue.edu}
\affiliation{Department of Physics and Astronomy, Purdue University, West Lafayette,
            Indiana 47907, USA}            
\date{\today}

%\author{D\'{a}vid Hvizdo\v{s}\thanksref{addr1}
%        \and
%        Roman \v{C}ur\'{\i}k\thanksref{addr1,e1}
%        \and
%        Chris H.~Greene\thanksref{addr2,addr3,e2}
%}

%\thankstext{e1}{e-mail: roman.curik@jh-inst.cas.cz}
%\thankstext{e2}{e-mail: chgreene@purdue.edu}

%\institute{J. Heyrovsk\'{y} Institute of Physical Chemistry, ASCR, Dolej\v{s}kova 3, 
%            18223 Prague, Czech Republic\label{addr1}
%          \and
%          Department of Physics and Astronomy, Purdue University, West Lafayette,
%            Indiana 47907, USA\label{addr2}
%            \and
%            Purdue Quantum Science and Engineering Institute, Purdue University, West Lafayette,
%            Indiana 47907, USA\label{addr3}
%}

%\date{Received: date / Accepted: date}
% The correct dates will be entered by the editor

\begin{abstract}
A numerically solvable two-dimensional (2D) model, employed by the authors to study the dissociative recombination of H$_2^+$ in the ungerade symmetry [Phys. Rev. A {\bf 98}, 062706 (2018)], is extended to describe the collision process in the gerade symmetry of H$_2$. In this symmetry the ionization and dissociation processes are driven primarily by the direct, curve-crossing mechanism. The model is represented by a set of three coupled electronic channels in 2D, in the space of $s,p,d$ partial waves of the colliding electron. We demonstrate that the Born\-/Oppenheimer properties of the H$_2$ molecule in the relevant range of internuclear distances can be described by such a model. The molecular rotational degrees of freedom are accounted for by the rotational frame transformation. The numerical solution of the model is discussed and the resulting rovibrationally inelastic and dissociative recombination cross sections are compared with the available data. 
\end{abstract}

\maketitle

\section{Introduction}

The starting point of nearly all molecular theory is the Born\-/Oppenheimer approximation, in which the nuclear motion is frozen and the electronic Schr\"odinger equation is solved for energies and/or scattering phaseshifts versus the internuclear distance $R$.  For an atom-atom inelastic collision at a low energy, the scattering matrix can be computed using the potential energy curves and nonadiabatic couplings between those potentials.  But for the class of processes such as dissociative recombination, which convert incident electron energy into dissociation of the molecule into atomic fragments, it becomes more challenging to visualize the process because the incident channel is in the electronic continuum and therefore has no potential curve in the usual sense.  The present study discusses a flexible model that can describe such processes, even in systems as rich and complex as the 
${\it gerade}$ 
symmetries of H$_2$ and the isotopologues.

It should be stressed at the outset that accurate theoretical methods have been developed over the past several decades, most notably by Jungen, Fano, Dill and coworkers
\cite{Fano_PRA_1970,Atabek_Dill_Jungen_PRL_1974,Jungen_Atabek_JCP_1977,Jungen_Dill_JCP_1980}, which have been remarkably successful in describing the low energy scattering processes that arise for the {\it ungerade} 
electronic parity of H$_2$. All these methods are built around a central idea, the rovibrational frame transformation (FT) \cite{Chang_Fano_1972} that assumes good accuracy of the Born\-/Oppenheimer approximation (BOA) in some finite, presumably small, electronic volume. Electron motion in the outer space, where the BOA is no longer valid, was then treated by the multi-channel quantum defect theory (MQDT) 
\cite{Seaton_RPP_1983}.

In the case of the {\it gerade} states of H$_2$, the success of the pure frame transformation theory was limited to calculations of vibronic-energy levels
\cite{Ross_Jungen_PRL_1987,Ross_Jungen_PRA_1994a,Ross_Jungen_PRA_1994b}. A description of the dissociative recombination process in the {\it gerade} channels of H$_2$ has developed along the MQDT quasidiabatic theory of Giusti 
\cite{Giusti_JPB_1980} that employs an explicit coupling between the dissociative and autoionizing states inside the reaction volume. The theory was successfully applied to study the DR of H$_2^+$ in various initial vibrational
\cite{Giusti-Suzor_PRA_1983,Nakashima_JCP_1987,Schneider_etal_H2_1991,Takagi_PS_2002} and rovibrational
\cite{Takagi_H2rot_JPB_1993,Schneider_JPB_1997,Waffeu_Schneider_PRA_2011,Motapon_Schneider_HDinel_PRA_2014}
states.

Our aim is to present an alternative theoretical tool to study and understand the dissociative recombination process, the numerically solvable two-dimensional (2D) model. The model has one electronic and one nuclear degree of freedom and it can be solved to high precision, without making any physically motivated approximations, by employing the 2D $R$-matrix approach 
\cite{Curik_HG_2DRmat_2018}. A single channel version of the 2D model was previously applied to study the DR process in H$_2^+$ for the ungerade symmetry
\cite{Hvizdos_etal_2018,Curik_HG_2DRmat_2018}, which is predominantly controlled by the indirect mechanism. The necessary extensions of the model to describe the direct mechanism of H$_2$ in the $^1\Sigma_g$ symmetry are presented in Sec.~\ref{sec-model}. Section~\ref{sec-BOA} demonstrates the realistic nature of the model by comparing several Born\-/Oppenheimer properties of the model with the available literature. Inelastic cross sections for the dissociative recombination and rovibrational excitation processes are discussed in Sec.~\ref{sec-DR} and Sec.~\ref{sec-inel}, respectively. Finally, Sec.~\ref{sec-concl} offers a summary and concluding remarks.

\section{\label{sec-model}Curve-crossing model of the singlet H$_2$ gerade system}
\subsection{2D model Hamiltonian}

While a single electronic partial wave was sufficient to previously model the indirect mechanism that is dominant in the $^1\Sigma_u$ symmetry of H$_2$
\cite{Hvizdos_etal_2018,Curik_HG_2DRmat_2018},
the direct mechanism present in the $^1\Sigma_g$ channels requires a coupling of three partial waves with electron orbital angular momenta $l=0,1,2$.
The two-dimensional Hamiltonian includes a coupling potential matrix $V_{ll'}(R,r)$ as
\begin{equation}
\label{eq-3x2D-Ham}
H_{ll'}(R,r)=\left[H_l^{\mathrm{n}}(R) + H_l^{\mathrm{e}}(r)\right]\delta_{ll'} + V_{ll'}(R,r)\; ,
\end{equation}
where $\delta_{ll'}$ is the Kronecker delta. The electronic Hamiltonian is defined by
\begin{equation}
H_l^{\mathrm{e}}(r) = - \frac{1}{2}\frac{\partial^2}{\partial r^2} - 
\frac{1}{r} + \frac{l(l+1)}{2r^2}\;,
\end{equation}
and the nuclear Hamiltonian is
\begin{equation}
H_l^{\mathrm{n}}\left(R\right) = 
- \frac{1}{2M}\frac{\partial^2}{\partial R^2} + V_l^{0}\left(R\right)\; .
\end{equation}
The ionic nuclear potentials $V_l^{0}(R)$ (taken from Madsen and Peek \cite{Peek_Madsen_1971}) are the target potential energy curves of the $1s\sigma_g$ state for $l=0,2$, and of the $2p\sigma_u$ state for $l=1$.

The potential $V_{ll'}(R,r)$ is comprised of three diagonal interaction terms $V_{00}$, $V_{11}$, $V_{22}$ and two off-diagonal coupling terms $V_{01}$, $V_{12}$. The direct coupling of the $s$- and $d$-wave channels ($V_{02}$) is neglected. All the interaction terms share the form
\begin{equation}
\label{eq-model-gerade-el}
	V_{ll'}(R,r) = \lambda_{ll'}(R) e^{-r^2/\omega^2}\; ,
\end{equation}
and $\omega = 2$ bohr is the same for all combinations of $l,l'$. For the diagonal interaction potentials, the nuclear part is chosen as a sum of two Gaussian curves
\begin{equation}
\label{eq-model-gerade-nuc-1}
	\lambda_{ll}(R) = A_{ll} e^{-\left(\frac{R-B_{ll}}{C_{ll}} \right)^2} + D_{ll} e^{-\left(\frac{R-E_{ll}}{F_{ll}} \right)^2}\; .
\end{equation}
The first coupling potential $V_{01}(R,r)$ is written as
\begin{equation}
\label{eq-model-gerade-nuc-2}
	\lambda_{01}(R) = A_{01} e^{-\left(\frac{R-B_{01}}{C_{01}} \right)^4}\; ,
\end{equation}
and the second coupling potential $V_{12}$ contains a single Gaussian
\begin{equation}
\label{eq-model-gerade-nuc-3}
	\lambda_{12}(R) = A_{12} e^{-\left(\frac{R-B_{12}}{C_{12}} \right)^2}\; .
\end{equation}

Table \ref{tab-model-gerade} contains all of the coefficients $A_{ll'}-F_{ll'}$. The input distances are measured in bohr radii and the output potential energy is in hartree units. The constants defining the interaction potentials were chosen such that the corresponding Born\-/Oppenheimer potential energy curves approximately describe the H$_2$ system in the $^1\Sigma_g$ symmetry. More details are given in Section~\ref{sec-BOA}.
\begin{table}[hb]
\centering
\begin{tabular}{cccccc}
%\begin{tabular}{l*{5}{d{-1}}}
 \toprule
 & $V_{00}$ & $V_{11}$ & $V_{22}$ & $V_{01}$ & $V_{12}$ \\
 \midrule
 $A_{ll'}$ & 0.350802 & -0.744042 & -1.13912 & -0.1960 & -0.4 \\
 $B_{ll'}$ & 4.53741  & 2.89011   & 8.50097  & 0       & 2.8  \\
 $C_{ll'}$ & 2.10017  & 3.05122   & 7.04271  & 5.52884 & 4.0  \\
 $D_{ll'}$ & 0.168061 & -0.327764 & -1.61629 & -       & -    \\
 $E_{ll'}$ & 2.71464  & 7.33829   & 21.1034  & -       & -    \\
 $F_{ll'}$ & 1.17950  & 3.40565   & 14.3257  & -       & -    \\
 \bottomrule
\end{tabular}
\caption{Table of coefficients for present model potential $V_{ll'}(R,r)$.}
\label{tab-model-gerade}
\end{table}

\subsection{Details of the solution}

The implementation of the 2D $R$-matrix method, its solution and computation of the inelastic cross sections is based on Ref.~\cite{Curik_HG_2DRmat_2018}. For present calculations we used the Wigner-Eisenbud form \cite{Wigner_Eisenbud_PR_1947,Robicheaux_1991}
\begin{equation}
R_{ij}(E) = \frac{1}{2} \sum_p \frac{\left(i|\psi_p\right)\left(\psi_p|j\right)}{E_p-E}\;,
\end{equation}
where the sum includes all the eigenstates $|\psi_p\rangle$ of the Hamiltonian (\ref{eq-3x2D-Ham}) extended %(and symmetrized inside the 2D box) 
to include the 2D Bloch operator \cite{Curik_HG_2DRmat_2018} which guarantees that the Hamiltonian is real and symmetric. The scalar product $\left( . | . \right)$ is carried out on the 2D surface and the surface channel functions $\left.|j\right)$ are represented by the cation vibrational functions on the electronic surface for $l=0,2$, whereas they become the lower-lying bound atomic electronic states on the nuclear surface \cite{Curik_HG_2DRmat_2018}. The channel functions on the electronic surface for $l=1$ are not employed as the present calculations are energetically confined below the cation's dissociation limit. Note that the $l=1$ partial wave is fully coupled with the $s$- and $d$-waves inside the box by the form of the Hamiltonian (\ref{eq-3x2D-Ham}).

The electronic box radius $r_0$ is chosen large enough so that the highest physical hydrogenic Rydberg state that can be produced in dissociation will fit inside, e.g. $r_0 > 2n_{max}^2$. The radius of $r_0 = 50$ bohr with 100 B-spline basis functions has proven to provide convergent results. 
Similarly, the dissociative coordinate box radius $R_0$ should be chosen to be large enough to fully contain any vibrational wavefunction that can be excited during the collision process.
From the shape of the potential curves shown in Fig.~\ref{fig-model-potcurves-WD} it is evident that the nuclear $R$-matrix radius $R_0$ needs to be larger than 30 bohr in order to connect the 2D solutions to the free asymptotic solutions for energies up to $n=3$ asymptote. The resulting large 2D box has led to a very demanding basis set size.

This technical problem can be alleviated by observing that the Born\-/Oppenheimer solutions become accurate at larger internuclear distances $R$.
Therefore, the symmetric version of the Hamiltonian plus Bloch operator (\ref{eq-3x2D-Ham}) was diagonalized instead in a smaller box with $R_0$ = 12 bohr (with 120 B-spline functions) and the box solutions on the nuclear surface $R_0$ were connected to the Born\-/Oppenheimer solutions for $R > R_0$. The corresponding nuclear components of these solutions were obtained using generalized quantum defect theory based on the Milne phase-amplitude technique described in Ref. \cite{Greene_Rau_Fano_1982}, Section IV. B. An adaptation of this technique for the present study is given in the Appendix.

\section{\label{sec-BOA}Adiabatic properties of the model}

\subsection{Born-Oppenheimer potential energy curves}

\begin{figure}[th]
	\centering
	\includegraphics[width=0.5\textwidth]{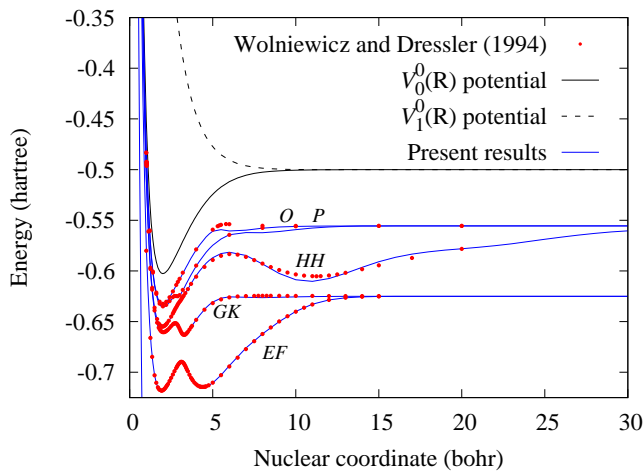}
	\caption{A comparison of the lowest second through sixth Born-Oppenheimer potential curves $E^{\mathrm{BO}}_k(R)$ obtained with the present model (blue curves) and those of Wolniewicz and Dressler (1994 )
	\cite{Wolniewicz_Dressler_JCP_1994} (points). 
	The black full and broken curves show the $V_l^{0}(R)$ of Madsen and Peek \cite{Peek_Madsen_1971}. Part of the lowest Born-Oppenheimer curve is also shown but it quickly falls below the range of the graph.}
	\label{fig-model-potcurves-WD}
\end{figure}

The model potential $V_{ll'}(R,r)$ was tailored to reproduce Born-Oppenheimer potential curves of excited $^1\Sigma_g^+$ states of the hydrogen molecule. Specifically, the $EF, GK$ and $H\bar{H}$ curves of L. Wolniewicz and K. Dressler (1994) \cite{Wolniewicz_Dressler_JCP_1994}. In the present model, these curves are computed by solving the $l$-coupled Born\-/Oppenheimer Schr\"{o}dinger equation
\begin{multline}
\label{eq-fixednuc-adiabatic}
\left[ H_l^{\mathrm{e}}(r) + V_l^{0}(R) - E^{\mathrm{BO}}_k(R)\right]\psi_{lk}(r;R) = \\
-\sum_{l'=0}^2 V_{ll'}(R,r) \psi_{l'k}(r;R)
\end{multline}
for a fixed $R$ and the electronic bound-state boundary condition $\psi_{lk}(\infty,R) \rightarrow 0$. A comparison can be seen in Fig. \ref{fig-model-potcurves-WD} which shows the present Born-Oppenheimer energies $E^{\mathrm{BO}}_k(R)$ and the $EF, GK$ and $H\bar{H}$ curves of L. Wolniewicz and K. Dressler (1994) \cite{Wolniewicz_Dressler_JCP_1985,Wolniewicz_Dressler_JCP_1994}. The $O$ and $P$ curves are also shown although they were not considered in the model potential $V_{ll'}(R,r)$ setup. The target cation Born-Oppenheimer curves $V^0_{l}(R)$ of Madsen and Peek  \cite{Peek_Madsen_1971} are also displayed as the fixed-$R$ ionization thresholds.

\subsection{Vibronic-energy levels}
The second test to explore the realistic nature of the present model are the vibronic-energy levels for this symmetry of the H$_2$ molecule. The computed curves $E^{\mathrm{BO}}_k(R)$ can be employed in the one-dimensional nuclear Schr\"{o}dinger equation
\begin{equation}
\left[ -\frac{1}{2 M}\frac{\partial^2}{\partial R^2} + E^{\mathrm{BO}}_k(R) + V^{\mathrm{cor}}_{k}(R) \right]\phi_{km}(R) = E^{\mathrm{vib}}_{km} \phi_{km}(R)\; ,
\label{eq-vibronic}
\end{equation}
where $V^{\mathrm{cor}}_{k}(R)$ are the diagonal second order BOA correction terms easily obtained during the solution of (\ref{eq-fixednuc-adiabatic}). 
Their form is
\begin{equation}
\label{eq-vibronic-cor}
V^{\mathrm{cor}}_{k}(R) = \sum_{l=0}^2 \frac{1}{2 M}\Big\langle \frac{\partial}{\partial R}\psi_{lk}(r;R) \Big| \frac{\partial}{\partial R}\psi_{lk}(r;R)\Big\rangle_r\; ,
\end{equation}
where $\langle\;.\;|\;.\;\rangle_r$ indicates an integration over the electronic coordinate only. The equation (\ref{eq-vibronic}) is solved with the boundary condition $\phi_{km}(R_0)=0$ (at a reasonably large $R_0$) with the index $k$ identifying the $EF, GK, H\bar{H}$ curves. The resulting energies are compared with experimental data for the real hydrogen molecule \cite{Ross_Jungen_PRA_1994b} in Table \ref{tab-gerade-viblevels}.
\begin{table}[ht]
\begin{center}
\begin{tabular}{rrrrr}
\toprule
\multicolumn{3}{c}{Vibronic} & \multicolumn{1}{c}{Vibronic} & \multicolumn{1}{c}{Comparison} \\
\multicolumn{3}{c}{state} & \multicolumn{1}{c}{energy } & \multicolumn{1}{c}{to Ref. \cite{Ross_Jungen_PRA_1994b} } \\
\multicolumn{3}{c}{ } & \multicolumn{1}{c}{(cm$^{-1}$)} & \multicolumn{1}{c}{(cm$^{-1}$)}\\
\midrule
E0   & & &   98949.68  &   215.10 \\
F0   & & &   99328.61  &    35.31 \\
F1   & & &  100500.70  &    58.23 \\
E1   & & &  101366.27  &   128.48 \\
F2   & & &  101642.30  &    56.63 \\
F3   & & &  102735.15  &    43.13 \\
E3   & & &  103485.05  &    74.54 \\
F4   & & &  103820.07  &    18.47 \\
F5   & & &  104719.16  &    11.45 \\
EF9  & & &  105387.11  &    -2.21 \\
EF10 & & &  106011.85  &   -45.69 \\
EF11 & & &  106768.25  &   -55.18 \\
EF12 & & &  107495.45  &   -69.58 \\
EF13 & & &  108186.65  &   -88.09 \\
EF14 & & &  108881.12  &   -87.57 \\
EF15 & & &  109573.13  &   -79.23 \\
EF16 & & &  110247.41  &   -84.03 \\
EF17 & & &  110903.03  &  -108.84 \\
 & GK0 & &  111381.77  &   247.04 \\
 & GK1 & &  111546.31  &   266.36 \\
EF19 & & &  112176.81  &   -70.72 \\
 & & H0  &  112782.32  &   175.25 \\
\bottomrule
\end{tabular}
\caption{\label{tab-gerade-viblevels}$EF$, $GK$ and $H\bar{H}$ curve vibronic energies. The first triple-column identifies the vibronic state with the first 9 states, using a single letter to denote whether they are situated within the first ($E$) or second ($F$) minimum of the $EF$ curve. The second column contains the energies relative to the $N=0,m=0$ level of $\tilde{X}$ $^1\Sigma_g^+$ ground state and the third column is the difference between our result and the observed value of Ref. \cite{Ross_Jungen_PRA_1994b} (observed - calculated).}
\end{center}
\end{table}

\vspace{-4mm}
\subsection{\label{ssec-res}Resonance $2p\sigma^2$}
The next adiabatic property tested for this model is the $2p\sigma^2$ resonance. This resonance exists for the energies of the neutral H$_2$ above the ionization threshold. Between the internuclear distances of 2 and 3 bohr it dives below the $V^0_0(R)$ threshold and produces a series of avoided crossings in the $^1\Sigma_g$ Rydberg levels, partially visible in Fig.~\ref{fig-model-potcurves-WD}.
In order to analyze the fixed-$R$ electronic continuum states of H$_2$ we extend the Hamiltonian in Eq.~(\ref{eq-fixednuc-adiabatic}) by adding the electronic Bloch operator (for simplicity $E^{\mathrm{BO}}_k$ is relabeled as $E_k$) 
\begin{multline}
\label{eq-fixednuc-adiabatic-Bloch}
\left[ H_l^{\mathrm{e}}(r) + V_l^{0}(R) - E_k(R) + \frac{1}{2} \delta(r-r_0)\frac{\partial}{\partial r}\right]\psi_{lk}(r;R) = \\
-\!\sum_{l'=0}^2 V_{ll'}(R,r) \psi_{l'k}(r;R)\; ,
\end{multline}
where the internuclear distance $R$ is a fixed parameter. The Born-Oppenheimer electronic $R$ matrix can be expanded over the poles in the Wigner-Eisenbud form
\begin{equation}
\label{eq-Rmat-33}
R_{ll'}(E,R) = \frac{1}{2}\sum_k
\frac{\psi_{lk}(r=r_0;R)\,\psi_{l'k}(r=r_0;R)}
{E_k(R) - E}\; ,
\end{equation}
with $r_0$ being the $R$-matrix radius and $E$ denotes the vertical energy. Let  $f_{l}(\epsilon_l,r)$, $g_{l}(\epsilon_l,r)$ stand for the regular and irregular Coulomb wave functions evaluated at $r=r_0$ and the energy $\epsilon_l=E-V^{0}_{l}(R)$. This independent pair of Coulomb solutions is identical to the $\{s,-c\}$ pair from Seaton's work \cite{Seaton_RPP_1983,Seaton_2002}. We denote their $r$-derivatives as $f'_l$, $g'_l$. The $R$ matrix can be transformed to the short-range $K$ matrix via the known relation \cite{Aymar_Greene_LKoenig_1996}
\begin{equation}
\label{eq-R2K-33}
\underline{K} = \left( \underline{f} -
\underline{f}' \underline{R} \right)
\left( \underline{g} -
\underline{g}' \underline{R} \right)^{-1}\; ,
\end{equation}
where underlined symbols denote matrices, in this case indexed by $ll'$. The $\underline{f}$, $\underline{g}$, $\underline{f}'$, $\underline{g}'$ are diagonal matrices. The short-range matrix $K_{ll'}(E,R)$ is computed for total energies above the $s$- and $d$- H$_2^+$ potential curve $V_{l=0,2}^{0}(R)$ and below the repulsive $p$-curve $V_{l=1}^{0}(R)$. This means that one of the three channels of this matrix is closed and the asymptotic component of all the three corresponding solutions are exponentially growing in this closed channel. The unphysical behavior is remedied by the MQDT technique called elimination of closed channels \cite{Seaton_RPP_1983,Aymar_Greene_LKoenig_1996}. In the present, simple three-channel case, the equation for the open-channel physical $K$ matrix at each $R$ can be written as
\begin{equation}
\label{eq-K-elim-33}
K_{ll'}^\mathrm{phys}(R) = K_{ll'}(R) - K_{l1}(R) \left[ K_{11}(R)+\mathrm{tan}\beta(R) \right]^{-1}K_{1l'}(R)\;,\;\;\;\; 
\end{equation}
where $l,l' = 0,2$, with
\begin{equation}
\nonumber
\beta(R) = \frac{\pi}{\sqrt{2\left(V_1^{0}(R)-E\right)}}\; .
\end{equation}
The explicit energy dependence is omitted in the above equations for clarity. 

\begin{figure}[thb]
	\centering
	\includegraphics[width=0.5\textwidth]{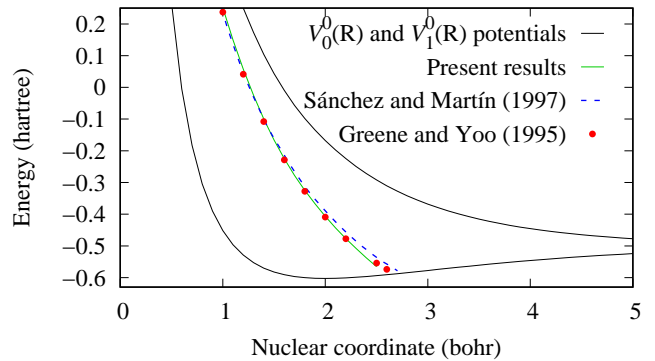}
	\caption{Comparison of resonance positions between the present model (full green curve) and calculations of S\'{a}nchez and Mart\'{i}n (1997) \cite{Sanchez_Martin_1997} (dashed blue curve) and Greene and Yoo (1995) \cite{Greene_Yoo_JPC95} (red points).}
	\label{fig-model-resonance}
\end{figure}

The eigenvalues of the physical 2$\times$2 $\underline{K}^\mathrm{phys}$ matrix are $\mathrm{tan}\left[\delta_{i}(E,R)\right]$, where the $\delta_{i}(E,R)$ are the Born-Oppenheimer physical eigenphase shifts.
Sampling these eigenphase shifts at a selected $R$ for a dense series of energies above the $V^0_{0}(R)$ curve results in step-like ascending curves. The steps in the sum of these curves show the positions of resonances.
These along with the resonance widths obtained from the energy derivative of $\delta_{i}(E,R)$ can be compared with accurate data \cite{Greene_Yoo_JPC95,Sanchez_Martin_1997} . The table \ref{tab-gerade-resonance} and Fig. \ref{fig-model-resonance} show the position $E_\mathrm{r}$ and width $\Gamma$ (only in the table) of the $2p\sigma^2$ resonance in H$_2$ for select values of $R$.

Note that the elimination of the closed $p$-channel is an important step in this study because the studied resonance is a Feshbach resonance described by the excited cation core $2p\sigma^1$ and the scattered electron in the closed channel with a wave function resembling the $2p\sigma^1$ orbital at small electronic distances.
\begin{table}[thb]
%\resizebox{0.5\textwidth}{!}{
\begin{tabular}{rrrrr}
\toprule
$R$ (bohr) & $E_\mathrm{r}$ (model) & $\Gamma$ (model) & $E_\mathrm{r}$ \cite{Greene_Yoo_JPC95} & $\Gamma$ \cite{Greene_Yoo_JPC95}\\
\midrule
1.0   & $ 0.25171$  &  0.028  &  $ 0.23726$  &  0.0229 \\
1.2   & $ 0.05753$  &  0.035  &  $ 0.04158$  &  0.0241 \\
1.4   & $-0.09948$  &  0.043  &  $-0.10799$  &  0.0251 \\
1.6   & $-0.22494$  &  0.052  &  $-0.22886$  &  0.0259 \\
1.8   & $-0.32475$  &  0.059  &  $-0.32768$  &  0.0265 \\
2.0   & $-0.40663$  &  0.068  &  $-0.40912$  &  0.0288 \\
2.2   & $-0.47684$  &  0.077  &  $-0.47731$  &  0.0335 \\
2.5   & $-0.56532$            & \multicolumn{1}{c}{-} &  $-0.55432$  &  0.0459 \\
2.6   & \multicolumn{1}{c}{-} & \multicolumn{1}{c}{-} &  $-0.57388$  &  0.0525 \\
\bottomrule
\end{tabular} %}
\caption{Positions $E_\mathrm{r}$ and widths $\Gamma$ of the $2p\sigma^2$ resonance in H$_2$ for the present model compared with Greene and Yoo (1995) \cite{Greene_Yoo_JPC95}. Results are given in atomic units. Our model resonance curve intersects the bottom potential curve in a slightly steeper manner, hence the undefined $E_\mathrm{r}$ and/or $\Gamma$ for the highest $R$ values.}
\label{tab-gerade-resonance}
\end{table}

\subsection{\label{ssec-qd}Born-Oppenheimer quantum defects}
The present model can also be used to compute quantum defects of the Born\-/Oppenheimer potential curves, and compare them to previously reported data \cite{Ross_Jungen_PRA_1994a}. 
The procedure is very similar to the one shown in the Sec.~\ref{ssec-res} up to the calculation of the short-range $K$ matrix in Eq.~(\ref{eq-R2K-33}). The pair of asymptotic functions $f_l$ and $g_l$ from the Sec.~\ref{ssec-res} is changed to a different pair $\{f^0,g^0\}$, denoted as $\{f,-h\}$ in Seaton's work \cite{Seaton_RPP_1983,Seaton_2002}. The new pair of the Coulomb functions has better analytic behavior for deep negative energies that occur in the present $p$-wave channel at short internuclear distances $R$. This different pair of asymptotic functions leads to a different $K$ matrix (denoted as $K^0$ here) with different phase-shift information (denoted as $\pi \eta$) with respect to these different regular and irregular Coulomb functions. 
\begin{figure}[th]
	\centering
	\includegraphics[width=0.5\textwidth]{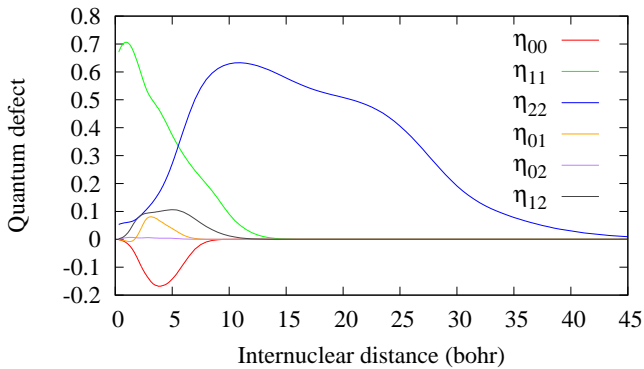}
	\caption{The $R$-dependence of the full quantum defect matrix at $E=V^0_{0}(R)$.}
	\label{fig-model-defectmat}
\end{figure}

Furthermore, the closed channel exponential growth will not yet be eliminated, as the short-range quantum defect matrix is defined to include both open and closed channels. The $K^0$ matrix can be diagonalized as follows
\begin{equation}
\label{eq-K-diag-33}
K^{0}_{ll'} = \sum_{i=0}^2 U_{li} \mathrm{tan}(\pi\eta_{i}) U_{l'i}\; .
\end{equation}
The quantum defect matrix $\eta_{ll’}$ can then be reconstructed through the same unitary transformation, i.e.
\begin{equation}
\label{eq-QDF-33}
\eta_{ll'} = \sum_{i=0}^2 U_{li} \eta_{i} U_{l'i}\; .
\end{equation}
\begin{figure}[tbh]
	\centering
	\includegraphics[width=0.5\textwidth]{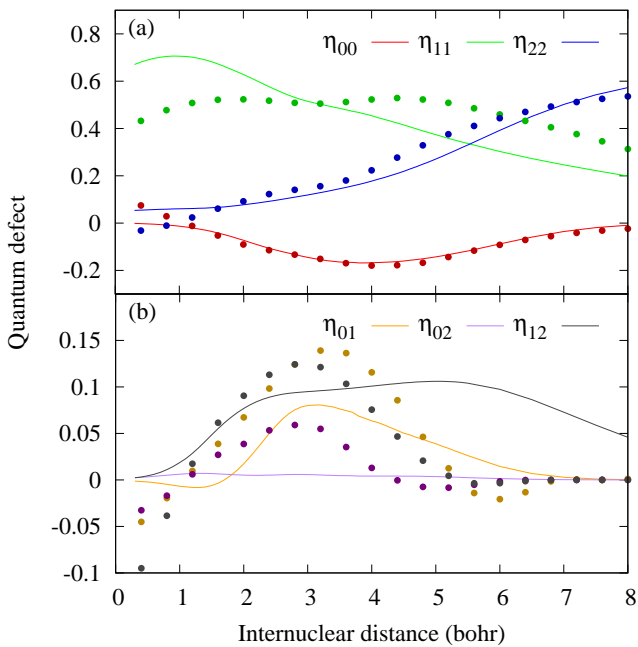}
	\caption{Zoomed-in comparison of the full quantum defect $\eta-$matrix elements (curves) with the data extracted from the work of Jungen and Ross \cite{Ross_Jungen_PRA_1994a} (points) ($R$-dependence at $E=V^0_{0}(R)$). Separated into (a) diagonal elements and (b) off-diagonal elements.}
	\label{fig-model-JRdefect}
\end{figure}
The dependence of the full $\eta$ matrix on the internuclear distance $R$ is displayed in Fig. \ref{fig-model-defectmat}. The short-range $R$-matrix radius was chosen at $r_0 = 8$ bohr and the electronic potential $V_{ll'}(R,r)$ could be neglected for $r>r_0$. The energy was fixed at the zero collision energy, i.e. total energy $E=V^0_{0}(R)$. Fig.~\ref{fig-model-JRdefect} compares the diagonal matrix elements $\eta_{ll}(E,R)$ to data extracted from Ref. \cite{Ross_Jungen_PRA_1994a}. Note that the quantum defects $\eta_{ll'}$ in Ref.~\cite{Ross_Jungen_PRA_1994a} are defined from the $K^0$ matrix on an element-by-element basis, i.e. $\eta_{ll'}^{[\rm Ref.8]} = \pi^{-1} \arctan K^0_{ll'}$, while in the present study a similar relation is used between the full $\eta$ and $K^0$ matrices. Therefore, in order to allow a comparison, the $K^0$ matrices from Ref.~\cite{Ross_Jungen_PRA_1994a} were converted to the $\eta$ matrices defined by equations (\ref{eq-K-diag-33}) and (\ref{eq-QDF-33}).  An electronic {\it ab initio} $R$-matrix calculation carried out by Bezzaouia, Jungen, and Telmini \cite{Jungen_Sigma_g_2004} also determined $R$-dependent $\eta$-matrices for this symmetry of H$_2$, but they adopted a larger channel set and for this reason the matrices from that calculation cannot be directly compared with our present results.

\section{\label{sec-DR}Dissociative recombination}

The DR cross section computed as the essentially exact solution of the model presented in this work can be compared with the available literature data. It is important to note that the present model includes only $^1\Sigma_g$ symmetry through which the DR process can proceed. The presented model does not include molecular rotational degrees of freedom. However, the rotational motion can be included with the rotational frame transformation. Discussion of this extension will be the subject of the second part of this section.

\subsection{Rotationless DR}
\begin{figure}[thb]
\begin{center}
\includegraphics[width=0.49\textwidth]{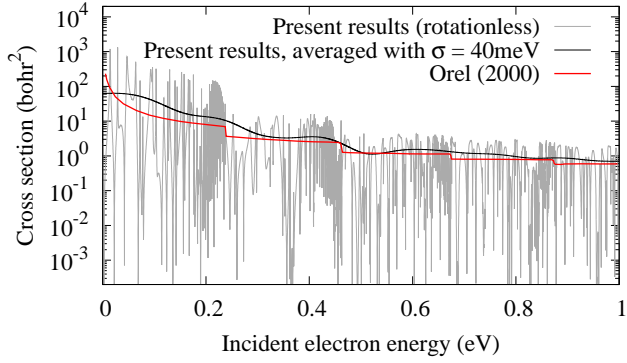}
\caption{\label{fig-dr_orel}
Comparison of the computed HD$^+$ DR cross section with the time-dependent simulations of A. Orel \cite{Orel_PRAR_2000} (red curve).
The grey curve shows our present results and the black curve is the same data averaged with a Gaussian distribution with $\sigma = 40$ meV.}
\end{center}
\end{figure}

Computed DR cross sections for the HD$^+$ cation are compared in Fig.~\ref{fig-dr_orel} with time-dependent calculations \cite{Orel_PRAR_2000}. In the previous calculations the wave-packet method solved a time-dependent Schr\"{o}dinger equation for nuclei with a local complex potential (LCP) that accounted for the autoionization process. The calculations were carried out in the $^1\Sigma_g$ symmetry for a single electronic partial wave that dominantly describes the diabatic neutral curve: the $d$ wave. 
The previous calculations \cite{Orel_PRAR_2000}, however, did not include any treatment of the closed electronic channels that belong to the vibrationally excited neutral states. These vibrational Feshbach resonances create the rich resonant structures converging to each of the excited cation thresholds.
\begin{figure}[th]
\begin{center}
\includegraphics[width=0.49\textwidth]{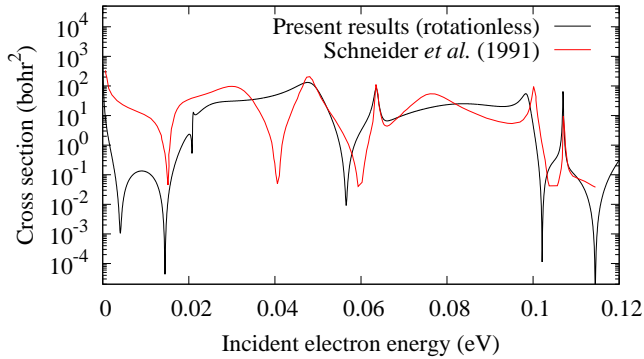}
\caption{\label{fig-dr_ifs}
Comparison of the computed H$_2^+$ DR cross section with calculations of I.F. Schneider \textit{et al.} (1991)  \cite{Schneider_etal_H2_1991}.
}
\end{center}
\end{figure}

In the following Fig.~\ref{fig-dr_ifs}, the previous calculations of Schneider \textit{et al.} (1991) \cite{Schneider_etal_H2_1991} were obtained using an MQDT treatment, and therefore the positions and presence of the closed-channel resonances can be compared. 
The calculations of \cite{Schneider_etal_H2_1991} were carried out in the $^1\Sigma_g$ symmetry with $R$-dependent $s$- and $d$-wave quantum defects.
It can be seen that apart from a single resonance, the two calculations do show a correlation. The resonance observed previously
\cite{Schneider_etal_H2_1991} around 40~meV is missing in the present calculations. Since it belongs to the $\nu$ = 6 threshold (with $n$ = 3) \cite{Schneider_etal_H2_1991}, its position is strongly sensitive to the inaccuracies of our potential energy curves shown in Fig.~\ref{fig-model-potcurves-WD}.

\subsection{DR with the rotational frame transformation}

The rotational degrees of freedom can be incorporated into the present calculations in the form of an approximation, the so-called rotational frame transformation (FT) \cite{Chang_Fano_1972}. The rotational FT is carried out on the short-range $K$ matrix (defined by Eq.~(44) in Ref.~\cite{Curik_HG_2DRmat_2018}) , before the elimination of closed channels. 
The rotational frame transformation transforms the body-frame set of quantum numbers
$(l,\Lambda)$ to the laboratory frame quantum numbers $(l,j)$ as
\begin{equation}
\label{eq-K-RFT}
K^{J}_{\nu j l,\nu' j' l'} = \sum_\Lambda U^{l J \eta}_{j \Lambda}\, K^\Lambda_{\nu l, \nu' l'}\, U^{l' J \eta}_{j'\Lambda} \;,
\end{equation}
where the unitary matrices $\underline{U}$ involve a Clebsch-Gordan coefficient, and can be found in Ref.~\cite{Chang_Fano_1972}. 
\begin{figure}[bht]
\begin{center}
\includegraphics[width=0.49\textwidth]{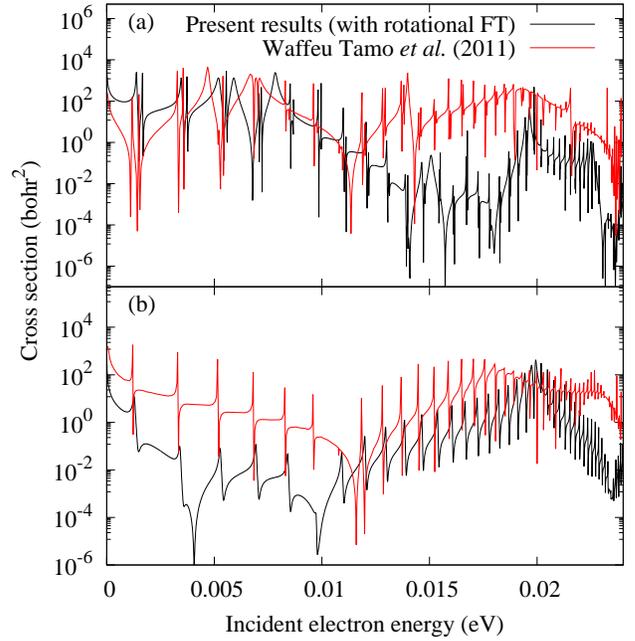}
\caption{\label{fig-dr_tamo}
Comparison of the computed HD$^+$ DR cross section with calculations of F.O. Waffeu Tamo \textit{et al.} (2011) \cite{Waffeu_Schneider_PRA_2011}. The initial rotational state is $j=1$ and the total angular momentum is (a) $J=3$ and (b) $J=1$.}
\end{center}
\end{figure}
Note that in the present $^1\Sigma_g$ model, the above sum over $\Lambda$ consists of a single term $\Lambda=0$ because we are neglecting the $\Pi$ state quantum defects of H$_2$ or HD. 
Furthermore, the parity quantum number $\eta$ is fixed to $\eta=1$, because $\eta=-1$ does not contribute to $\Lambda=0$.
The short-range $K$ matrix (\ref{eq-K-RFT}) then undergoes the MQDT procedure of elimination of closed channels defined by $(\nu,j)$. Closed-channel resonances now converge to each of these rovibratonal channel thresholds. An example of this behaviour together with a comparison with calculations of F.O. Waffeu Tamo \textit{et al.} (2011) \cite{Waffeu_Schneider_PRA_2011} is shown in Fig.~\ref{fig-dr_tamo}. The initial rovibrational state of the cation is $(0,1)$ and the cross section is shown for the total $J=1$ and $J=3$ symmetries. 
\begin{figure}[th]
\begin{center}
\includegraphics[width=0.49\textwidth]{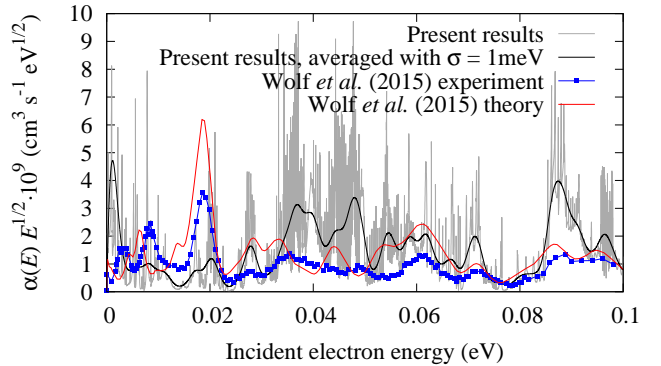}
\caption{\label{fig-dr_wolf}
Reduced rate coefficient $\alpha E^{1/2}$ for the DR of HD$^+$. Present averaged results (thick black curve) are compared with the experimental data (blue dots connected with line) 
\cite{Wolf_2015}
and the theory (red curve) \cite{Wolf_2015}. Contributions of the initial rotational levels are weighted by the Boltzmann distribution at 300K.
}
\end{center}
\end{figure}

Present cross sections are somewhat small at lower energies due to the closed-channel resonance already seen in the rotationless results in Fig.~\ref{fig-dr_ifs} around 4 meV. The second visible difference is a small shift of the closed-channel resonances converging to the $(\nu,j)=(0,3)$ threshold. This shift of a few cm$^{-1}$ can be explained by the limited accuracy of the present model's setup.

The last comparison of the computed DR rates also includes the experimental data. The energy dependence of the reduced DR rate coefficient $\alpha(E) E^{1/2}$ is shown in Fig.~\ref{fig-dr_wolf} on the linear scale. In order to simulate the experimental conditions, the DR rate coefficients with the initial rotational states up to $j=5$ were averaged with the thermal Boltzmann distribution at 300K. Although many of the details in the DR rate results disappear in the thermal average, this comparison is an important test for the absolute values of the computed rates. Fig.~\ref{fig-dr_wolf} shows that while the present DR rate has correct order of magnitude, some of of the higher-level resonant structures are overestimated by factor of 2-3 by the present calculations and the peaks under 20 meV are weaker in our results.

\section{\label{sec-inel}Rovibrational excitation}
%%%%%%%%%%%%%%%%%%%%%%%%%%%%%%%%%%

It is important to note here that any comparison with the results for the rovibrationally inelastic processes will strongly suffer, due to the use of the single symmetry $^1\Sigma_g$ in the present model. While it is known that the dissociative recombination of H$_2^+$ is strongly governed by the direct mechanism present in the $^1\Sigma_g$ symmetry
\cite{Orel_PRAR_2000,Schneider_etal_H2_1991,Takagi_H2rot_JPB_1993},
the rotational and vibrational excitation generally requires a number of dominant symmetries for the convergence
\cite{Motapon_Schneider_HDinel_PRA_2014}.

\subsection{Vibrational excitation}

To make our vibrationally inelastic cross section as complete as possible, the present $^1\Sigma_g$ results are combined with previously published
\cite{Curik_HG_2DRmat_2018} results obtained from a simplified model of the ungerade symmetry. The first simplification rests in the use of the sole $^1\Sigma_u$ angular momentum projection, while the quantum defect of $\Pi$ states is not negligible \cite{Jungen_Atabek_JCP_1977}. 
\begin{figure}[h]
\begin{center}
\includegraphics[width=0.49\textwidth]{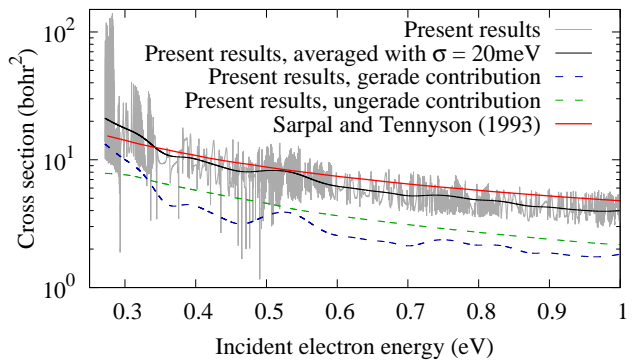}
\caption{\label{fig-ve_sarpal}
Vibrationally inelastic cross sections for H$_2^+$ $(\nu,j)$ transition $(0,0) \rightarrow (1,0)$. The broken lines show respective gerade and ungerade contributions to the total cross section depicted by the thick black curve. All the three curves are a result of the Gaussian distribution average with $\sigma = 20$ meV. Data of Sarpal and Tennyson 1993 \cite{Sarpal_Tennyson_MNRAS_1993} are displayed by the red curve.
}
\end{center}
\end{figure}
The second difference from the present study is the lack of the rotational degrees of freedom in the previous calculations. 

The results shown in Fig.~\ref{fig-ve_sarpal} indicate that the ungerade and gerade contributions to the vibrational excitation process are of a similar size. The resulting sum of the two contribution cross sections is compared with the ab-initio calculations of Sarpal and Tennyson \cite{Sarpal_Tennyson_MNRAS_1993}. The authors employed the adiabatic nuclei approximation of D.M. Chase
\cite{Chase_1956}
to compute the vibrationally inelastic cross section. In this approximation the physical, open-channel, scattering $T$ matrix is averaged over the initial and final vibrational states and thus this procedure can not produce the closed-channel resonances as seen in Fig.~\ref{fig-ve_sarpal}. For comparison present (and previous \cite{Curik_HG_2DRmat_2018} ungerade) results were convolved over a Gaussian distribution with the variance width of 20 meV.

\subsection{Rotational excitation}

\begin{figure}[th]
\begin{center}
\includegraphics[width=0.49\textwidth]{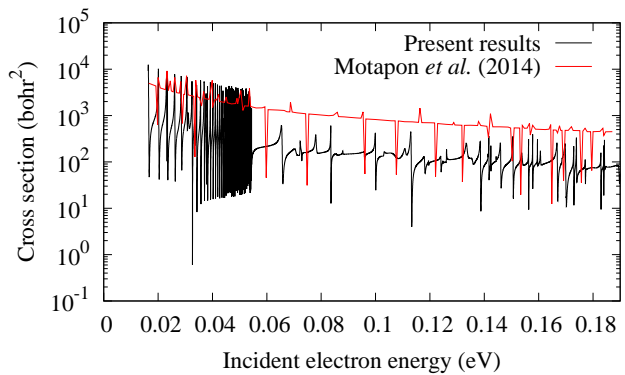}
\caption{\label{fig-re_motapon}
Rotationally inelastic cross sections for HD$^+$ $(\nu,j)$ transition $(0,0) \rightarrow (0,2)$. Results of the present mode are shown by the black curve, while the data taken from of Motapon \textit{et al.} (2014) \cite{Motapon_Schneider_HDinel_PRA_2014} are displayed by red line.
}
\end{center}
\end{figure}

A comparison of our present results for the rotational transition $0 \rightarrow 2$ of HD$^+$ with the calculations of Motapon \textit{et al.} (2014) 
\cite{Motapon_Schneider_HDinel_PRA_2014}
is shown in Fig.~\ref{fig-re_motapon}. Previous calculations 
\cite{Motapon_Schneider_HDinel_PRA_2014}
were carried out for all the relevant symmetry components, including $\Sigma$, $\Pi$, $\Delta$ states in gerade and ungerade symmetries with singlet and triplet spin components. On average the $^1\Sigma_g$ cross section contributes only 15\% of the total excitation probability, as can be seen in Fig.~\ref{fig-re_motapon}.

\section{\label{sec-concl}Conclusions}

The two-dimensional reactive model, that was developed to describe a collision between the electron and the H$_2^+$ cation in the ungerade symmetry 
\cite{Curik_HG_2DRmat_2018},
is extended in the present study to treat processes that involve the direct dissociative recombination mechanism. This mechanism is present in the gerade symmetry of the $e^- + \mathrm{H}_2^+$ collisions and this system has served as a prototype test of the 2D model. The present model was set up to approximately reproduce the Born-Oppenheimer $EF$, $GK$, and $H\bar{H}$ $^1\Sigma_g$ Rydberg levels of H$_2$. It was shown that all the other adiabatic properties, namely the resonance position and width at positive energies, the $\eta$ quantum defects and the vibronic levels, agreed qualitatively and to some extent also quantitatively with the available H$_2$ data.  In principle, our model could be readily extended to reproduce the known Born-Oppenheimer potential curves and fixed-nuclei resonance properties to arbitrary precision, when higher quantitative accuracy is desired.

The rotational degrees of freedom have been introduced in this study through the rotational frame transformation
\cite{Chang_Fano_1972} that was applied to the short-range $K$ matrix obtained by solving the $l$-coupled 2D model. This procedure allowed for modeling of rovibrational excitations as well as rovibrationally resolved dissociative recombination.
In case of the vibrational excitation process, our results demonstrate that the gerade and ungerade contributions are of a similar size and their sum agrees with the adiabatic calculations of Sarpal and Tennyson (1993) 
\cite{Sarpal_Tennyson_MNRAS_1993}. On the other hand, the present $^1\Sigma_g$ calculations for the rotational excitation $0 \rightarrow 2$ amount to only about 15\% of the excitation probability above the $j=2$ rotational threshold. This deficiency is caused primarily by contributions from the other symmetries (ungerade, spin triplets, non-$\Sigma$ projections of the angular momentum) that are not present in our calculations presented here.

In the case of dissociative recombination, the model was tested against the available rotationless and rotationally resolved calculations. In both cases, good correspondence for the closed-channel resonances was found except for one deeply closed-channel resonance that was found at lower energies here. We believe that this resonance shift is responsible for some of the quantitative disagreement with previous calculations
\cite{Schneider_etal_H2_1991,Waffeu_Schneider_PRA_2011}.

We stress that a precise tuning of the 2D model to the most accurate H$_2$ Born-Oppenheimer data has not been the aim of the present study. Our goal is instead to demonstrate that the model can incorporate all of the relevant physics contained in the $e^- + \mathrm{H}_2^+$ collisions driven by the direct (and previously indirect \cite{Hvizdos_etal_2018,Curik_HG_2DRmat_2018}) mechanism. As such it can serve as a physically relevant benchmark test for developing and testing those approximate theories that aspire to describe such a challenging non-adiabatic process as dissociative recombination.

\begin{acknowledgements}
The work of R\v{C} abd DH was supported by the Czech Science Foundation (Grant No. GACR 21-12598S).  CHG is supported in part by the U.S. Department of Energy, Office of Science, Basic Energy Sciences, under Award No. DE-SC0010545.
\end{acknowledgements}

\appendix

\section{Asymptotic functions from the Milne generalized QDT method}
After shrinking the 2D box to $R_0=12$ bohr the nuclear fragmentation surface channel functions $\left.|i_\mathrm{n}\right)$ have eigenenergies $E^{\mathrm{BO}}_{i_\mathrm{n}}(R_0)$ and each is connected to one of the Born-Oppenheimer potential curves (Fig. \ref{fig-model-potcurves-WD}).
The functions connecting these channels to the asymptotic region (taking the role of the Bessel functions in Ref. \cite{Curik_HG_2DRmat_2018}) are the solutions of
\begin{equation}
\label{eq-Milne-adi1}
\left[ \frac{\partial^2}{\partial R^2} + 2 M \left(E-E^{\mathrm{BO}}_{i_\mathrm{n}}(R)\right)\right] y_{i_\mathrm{n}}(R,E) = 0\;, \hbox{ for }R\geq R_0\; .
\end{equation}
Here we employ the Milne phase amplitude method described in section IV. B of Ref. \cite{Greene_Rau_Fano_1982}. It is known that {\it all} independent solutions to (\ref{eq-Milne-adi1}) can be constructed from any particular solution of the nonlinear Milne amplitude differential equation,
\begin{equation}
\label{eq-Milne-adi2}
\left[ \frac{\partial^2}{\partial R^2} + 2 M \left(E-E^{\mathrm{BO}}_{i_\mathrm{n}}(R)\right)\right] \alpha_{i_\mathrm{n}}(R,E) = \frac{1}{\alpha^3_{i_\mathrm{n}}(R,E)}\; ,
\end{equation}
via
\begin{equation}
\label{eq-Milne-general}
y_{i_\mathrm{n}}(R,E) = a \alpha_{i_\mathrm{n}}(R,E)
\mathrm{sin}\left(\int^{R} \alpha^{-2}_{i_\mathrm{n}}(R',E) dR' + b \right)\; ,
\end{equation}
with arbitrarily chosen constants $a$ and $b$. Following Ref. \cite{Greene_Rau_Fano_1982} we use the WKB-motivated boundary condition
\begin{equation}
\alpha_{i_\mathrm{n}}(R_0,E)=[2M(E-E^{\mathrm{BO}}_{i_\mathrm{n}}(R_0))]^{-1/4}\; ,
\end{equation}
and define the phase
\begin{equation}
\label{eq-Milne-phase}
\theta_{i_\mathrm{n}}(R,E)=\int^{R}_{R_0} \alpha^{-2}_{i_\mathrm{n}}(R',E) dR'\; ,
\end{equation}
to construct the pair
\begin{eqnarray}
\label{eq-Milne-F0}
\tilde{F}_{i_\mathrm{n}}(R) &=& \phantom{-}(2M/\pi)^{1/2}
\alpha_{i_\mathrm{n}} \left(R,E\right)
\sin\left[ \theta_{i_\mathrm{n}} \left(R,E\right) \right]\; , \\
\label{eq-Milne-G0}
\tilde{G}_{i_\mathrm{n}}(R) &=& -(2M/\pi)^{1/2} 
\alpha_{i_\mathrm{n}} \left(R,E\right)
\cos\left[ \theta_{i_\mathrm{n}} \left(R,E\right) \right]\; ,
\end{eqnarray}
which replaces the previous pair of Bessel functions $F^0_{i_\mathrm{n}}(R)$ and $G^0_{i_\mathrm{n}}(R)$ used in Ref. \cite{Curik_HG_2DRmat_2018}.
When using this pair in a weakly-closed channel we now need to alter the corresponding elements of the channel elimination procedure.
Let $R_1$ represent the distance at which our Born-Oppenheimer curves $E^{\mathrm{BO}}_{i_\mathrm{n}}(R_1)$ arrive very close to their asymptotic limits with a negligible error. We can define
\begin{equation}
\label{eq-Milne-phase-2}
\Theta_{i_\mathrm{n}}(E)=\int^{R_1}_{R_0} \alpha^{-2}_{i_\mathrm{n}}(R',E) dR'\; ,
\end{equation}
which is the additional phase shift that $\tilde{F}_{i_\mathrm{n}}(R),\tilde{G}_{i_\mathrm{n}}(R)$ accumulate before becoming a linear combination of spherical Bessel functions. Using the notation from Ref. \cite{Curik_HG_2DRmat_2018} we partially redefine the $\Gamma_{i\gamma}$ and $\Lambda_{i\gamma}$ matrices utilized in the eigenchannel form of the MQDT channel elimination equations as:
\begin{equation}
\label{eq-Milne-chanelim-G}
\Gamma_{i\gamma} = \left\{
\begin{array}{ll}
%U_{j\gamma} \sin(\beta_j + \pi\tau_\gamma)\; , & j \in Q_{\mathrm{e}} \\
U_{i\gamma} \sin(\Theta_i+ \pi\tau_\gamma)\; , &
i \in Q_{\mathrm{n}} \\
%\\
%U_{j\gamma} \sin\pi\tau_\gamma\; , & j \in P_{\mathrm{e}} \\
U_{i\gamma} \sin\pi\tau_\gamma\; , & i \in P_{\mathrm{n}} 
\end{array}
\right. ,
\end{equation}
\begin{equation}
\label{eq-Milne-chanelim-L}
\Lambda_{i\gamma} = \left\{
\begin{array}{ll}
%0\; , & j \in Q_{\mathrm{e}} \\
0\; , & i \in Q_{\mathrm{n}} \\
%\\
%U_{j\gamma} \cos\pi\tau_\gamma\; , & j \in P_{\mathrm{e}} \\
U_{i\gamma} \cos\pi\tau_\gamma\; , & i \in P_{\mathrm{n}} 
\end{array}
\right. ,
\end{equation}
in the nuclear fragmentation channels. The rest of the channel elimination is carried out the same as before; however, note that now, with the rotational FT approach, the channel index $i$ represents a combination of indices $(\nu,j,l)$ on the electronic fragmentation surface. Solution of the generalized eigenvalue problem
\begin{equation}
\label{eq-chanelim-singular}
\underline{\Gamma}\, \underline{A} = \underline{\Lambda}\, \underline{A}
\tan \underline{\delta}\; ,
%\sum_{\gamma} \Gamma_{i\gamma} A_{\gamma\rho} = \sum_{\gamma} \Lambda_{i\gamma} A_{\gamma\rho} \tan %\delta_{\rho}\; ,
\end{equation}
for the unknown $A_{\gamma\rho}$ and $\delta_\rho$ gives us the transformation matrix into the physically open set of collision eigenchannels $T_{i \rho}$ (see Ref. \cite{Curik_HG_2DRmat_2018}).
Then $T_{i \rho}$ is used to construct the open channel reaction matrix
\begin{equation}
\label{eq-2drm-Kmilne}
K_{ii'} = \sum_\rho T_{i\rho} \mathrm{tan}\left(\delta_{\rho}\right) T_{i'\rho}\; ,\quad
i,i' \in P_{\mathrm{e}} \cup P_{\mathrm{n}}\; .
\end{equation}
This is not yet the physical reaction matrix in the nuclear fragmentation channels, because the base pair $\{F,G\}$ does not have energy-normalized amplitudes, and because their phases are connected to Milne functions instead of to Bessel functions. Asymptotically, the Milne functions behave as a simple linear combination of Bessel functions characterized by two sets of coefficients
\begin{eqnarray}
\label{eq-MilneBess-F}
\tilde{F}_{i_\mathrm{n}}(R) &=& F_{i_\mathrm{n}}(R) B^{-1/2}_{i_\mathrm{n}} \; , \\
\label{eq-MilneBess-G}
\tilde{G}_{i_\mathrm{n}}(R) &=& G_{i_\mathrm{n}}(R) B^{-1/2}_{i_\mathrm{n}} - F_{i_\mathrm{n}}(R) B^{-1/2}_{i_\mathrm{n}} \mathcal{G}_{i_\mathrm{n}} \; , \\
\nonumber
 \mathrm{for }\; R &\rightarrow& \infty \; ,
\end{eqnarray}
where the energy-normalized pair $F_{i_\mathrm{n}}(R),G_{i_\mathrm{n}}(R)$ contain a phase shift and amplitude revision compared to $F^0_{i_\mathrm{n}}(R),G^0_{i_\mathrm{n}}(R)$. While $F_{i_\mathrm{n}}(R),G_{i_\mathrm{n}}(R)$ are rotated in phase compared to the usual spherical Bessel solutions, removing this phase shift is not necessary if we do not wish to compute the elastic scattering cross sections or any differential angle-dependent observables.
The coefficients $B_{i_\mathrm{n}}$ and $\mathcal{G}_{i_\mathrm{n}}$ can be computed numerically when constructing the Milne solutions. Additionally defining $B_{i_\mathrm{e}}=1$ and $\mathcal{G}_{i_\mathrm{e}}=0$ in the ionization channels allows us to transform the open channel reaction matrix into the physical reaction matrix simply via the matrix equation
\begin{equation}
\label{eq-2drm-Ktransform}
\underline{K}^{\mathrm{phys}} = \underline{B}^{1/2} (\underline{K}^{-1}+\underline{\mathcal{G}})^{-1} \underline{B}^{1/2}.
\end{equation}
This $K$ matrix is then Cayley\-/transformed into the scattering matrix from which cross sections are computed.

\bibliographystyle{apsrev}
\bibliography{bibliography}

\end{document}